\documentclass[11pt,a4paper]{article}
\pdfoutput=1
\usepackage{jcappub}
\usepackage{slashed}
\usepackage{stackengine}
\def\be{\begin{equation}}
\def\ee{\end{equation}}
\def\bea{\begin{eqnarray}}
\def\eea{\end{eqnarray}}

\def\ma{\mathfrak{a}}
\def\mg{\mathfrak{g}}
\def\mm{\mathfrak{m}}
\def\mn{\mathfrak{n}}
\def\me{\mathfrak{u}}

\def\mv{\mathfrak{v}}
\def\ms{\mathfrak{s}}

\def\mf{\mathfrak{f}}
\def\mJ{\mathfrak{J}}

\def\mR{\mathfrak{R}}
\def\mL{\mathfrak{L}}
\def\td{\text{d}}

\begin{document}

\title{Covariant Information Theory and Emergent Gravity} 

\author{Vitaly Vanchurin}

\emailAdd{vvanchur@d.umn.edu}

\date{\today}

\affiliation{Department of Physics, University of Minnesota, Duluth, Minnesota, 55812 \\
Duluth Institute for Advanced Study, Duluth, Minnesota, 55804}

\abstract{Informational dependence between statistical or quantum subsystems can be described with Fisher information matrix or Fubini-Study metric obtained from variations/shifts of the sample/configuration space coordinates. Using these (non-covariant) objects as macroscopic constraints we consider statistical ensembles over the space of classical probability distributions (i.e. in statistical space) or quantum wave-functions (i.e. in Hilbert space). The ensembles are covariantized using dual field theories with either complex scalar field (identified with complex wave-functions) or real scalar field (identified with square roots of probabilities). We construct space-time ensembles for which an approximate Schrodinger dynamics is satisfied by the dual field (which we call {\it infoton} due to its informational origin) and argue that a full space-time covariance on the field theory side is dual to local computations on the information theory side. We define a fully covariant information-computation tensor and show that it must satisfy certain conservation equations. 

Then we switch to a thermodynamic description of the quantum/statistical systems and argue that the (inverse of) space-time metric tensor is a conjugate thermodynamic variable to the ensemble-averaged information-computation tensor. In (local) equilibrium the entropy production vanishes, and the metric is not dynamical, but away from the equilibrium the entropy production gives rise to an emergent dynamics of the metric. This dynamics can be described approximately by expanding the entropy production into products of generalized forces (derivatives of metric) and conjugate fluxes. Near equilibrium these fluxes are given by an Onsager tensor contracted with generalized forces and on the grounds of time-reversal symmetry the Onsager tensor is expected to be symmetric. We show that a particularly simple and highly symmetric form of the Onsager tensor gives rise to the Einstein-Hilbert term. This proves that general relativity is equivalent to a theory of non-equilibrium (thermo)dynamics of the metric, but the theory is expected to break down far away from equilibrium where the symmetries of the Onsager tensor are to be broken.

 }

\maketitle

\section{Introduction}

Quantum gravity is an ongoing attempt to unify quantum mechanics and general relativity within a single theory. There are many approaches to unification that had been proposed, but perhaps the most ambitious of all is to start with nothing but quantum mechanics and then to derive everything from it, i.e. space-time, Lorentz covariance, general relativity, corrections, etc. In all such attempts the first step should be a derivation of space-time from a state vector or a density matrix in a Hilbert space (which describes a state of the system) and a Hermitian matrix or a Hamiltonian operator (which describes dynamics of the system). In the recent years a number of interesting proposal for deriving space were put forward, which includes quantum entanglement \cite{VanRaamsdonk:2010pw, Swingle:2014uza},  tensor networks \cite{Swingle}, multi-dimensional scaling \cite{Carroll}  and information graph flow \cite{graph_flow}, but it is less clear how to derive space-time (see however Ref. \cite{Noorbala} and \cite{graph_flow}) or how general relativity might emerge. In this paper we will try to answer some of these questions in context of either classical or quantum theories of information. We will show that in certain limits general relativity is equivalent to a theory of non-equilibrium (thermo)dynamics of information. Note that the idea of viewing general relativity not as a fundamental phenomena, but as an emergent thermodynamic phenomena is not new. The first attempt to derive Einstein equations from the laws of thermodynamics was made in Ref. \cite{Jacobson} and more recently in Ref. \cite{Verlinde}. 

But before we dive into the questions of quantum gravity we will first construct a covariant information theory (which in a physicist's jargon should mean a ``second quantization'' of information.) In a sense we will be describing (classical or commutative) probability distributions over either classical probabilities (also known as second-order probabilities) or quantum states, but to make the notion more precise we will define statistical ensembles using dual fields which we shall call {\it infoton}. The reason why our analysis is directly related to the theories of information (as opposed to just probability theory) is that  the macroscopic constraint that we impose is a generalization and covariantization of the Fisher information metric and Fubini-Study metric. In fact, we would like to stress that our entire approach to covariantization of information and also to (perhaps un-)quantization of gravity is based on an explicit assumption that most of the microscopic parameters, which specify either classical probabilities or quantum states, are not observable, and one can only keep track of a very small subset of macroscopic parameters. And these parameters (at least in this paper) are taken to describe informational dependence (e.g. mutual information or quantum entanglements) between subsystems.

To illustrate the main idea, consider a quantum system of $N$ qubits whose {\it pure quantum states} are represented by rays, i.e. vectors in a $2^N$-complex dimensional Hilbert space normalized to unit length modulo an arbitrary phase factor. (It will be convenient to borrow some terminology from classical mechanics and to think of the Hilbert space as a phase space, the normalization as a constraint and of the overall phase as a symmetry.) More generally, one can consider a probability distribution over pure states or what is usually called a {\it mixed quantum states}, but might also be called a statistical ensemble of states in the Hilbert space. The mixed states are often described in terms of density matrices, but, as we shall argue below, in our case it is more convenient to describe such ensembles using partition functions of certain dual field theories. The main reason why the tools of statistical mechanics can be useful is that the Hilbert space already comes with a symplectic form (which is nothing but imaginary part of the inner product  \cite{Ashtekar}) and thus one can define a volume form, carry on integrations, define statistical ensembles, etc. 

What is, however, new in the Hilbert space, in contrast to the classical phase space, is that it also comes with a metric tensor, which is nothing but a real part of the inner product  \cite{Ashtekar}. Since all of the vectors that represent physical states are unit vectors, or points on a unit sphere,  these points are within distance of $O(1)$ from each other if we were allowed to move along geodesics. Now imagine that we are only allowed to move in $O(N^2)$ out of all  $O(2^N)$ possible directions in the Hilbert space. More precisely, we are allowed to only apply $O(N)$ of one-qubit gates or $O(N^2)$ of two-qubit gates. Then the relevant question is: what is the shortest distance (or a relative computational complexity) between an arbitrary pair of states? This is like playing a very high-dimensional maze (which we previously called {\it the quantum maze} \cite{Vanchurin}) with a lot of walls and a very few pathways. In such a geometric view the Hilbert space of qubits is equipped not only with the inner product metric \cite{Ashtekar} but also with what one call a computational metric (see Ref. \cite{Nielsen} for details). In fact the computational metric can be used to define statistical ensembles of the Hilbert space trajectories. For example, in Ref. \cite{Vanchurin} we constructed an ensemble which suppresses trajectories that cannot be obtained by consecutive applications of either one- or two-qubit gates. Such ensembles are defined using a dual scalar field which takes values in both space and time.

In this paper we will also apply the dual description for constructing (more general) ensembles of quantum/statistical states, but this time subject to macroscopic constraints such as the mutual information, the quantum entanglements or the amount of computations. Moreover, once the statistical description is established we will seek for a thermodynamic framework in which the macroscopic variables will be treated as thermodynamic quantities. In Table \eqref{Table} we summarize a mapping between information theory, field theory and thermodynamic variables that we shall use throughout the paper
{{\renewcommand{\arraystretch}{1.5}\begin{table}
\begin{center}
\begin{tabular}{ |c|c|c|} 
 \hline
 Information Theory & Field Theory & Thermodynamics  \\ 
  \hline
   \hline
sample/configuration space, $\vec{x}$ & physical space, $\vec{x}$  &  physical space, $\vec{x}$ \\
 \hline
 state vector  $|\psi\rangle$ or $\vec{p}$ & &\\
  \hline
square root of prob., $\sqrt{p(\vec{x})}$ & real scalar field, $\varphi$ &   \\ 
 \hline
quantum wave-function, $\psi(\vec{x})$ & complex scalar field, $\varphi$ &  \\ 
 \hline
 probability density, $\psi(\vec{x})^*\psi(\vec{x})$ & probability scalar, $\varphi^*\varphi$  & particle number, $\mn$    \\ 
  \hline
information matrix, $A_{ij}$  & information tensor, ${\cal A}_{ij}$ &   information tensor, $\ma_{ij}$   \\
 \hline
 & mass squared, $\lambda$ & chemical potential, $\mm$ \\
  \hline
 & spatial metric, $g_{ij}$ & spatial metric, $\mg_{ij}$ \\
  \hline
  & free energy, $- \hbar \log({\cal Z})$ & free energy, $\mf$ \\
  \hline
 &information-computation  & information-computation \\ 
& tensor, ${\cal A}_{\mu\nu}$ &  tensor, $\ma_{\mu\nu}$ \\
 \hline
  & space-time metric, $g_{\mu\nu}$ &  space-time metric, $\mg_{\mu\nu}$ \\
    \hline
 & & entropy production, $\frac{1}{2\kappa} {\mR}$ \\
   \hline
  & & generalized forces,  $\mg_{\alpha\beta,\mu}$\\
    \hline
  & & Onsager tensor, $\mL^{\mu\nu\;\alpha\beta\;\gamma\delta} $\\
 \hline
\end{tabular} 
\end{center}
 \label{Table}
\caption{Mapping between information theory, field theory and thermodynamic variables. }
\end{table}}. 
Note that, on one hand the more microscopic quantities such as the state vectors ($|\psi\rangle$ or $\vec{p}$ ) have no analogs in a thermodynamic description of the system, and on the other hand the more macroscopic quantities such as the non-equilibrium entropy production  $\frac{1}{2\kappa} {\mR}$ is a lot more  difficult (although not impossible) to describe on the information theory side.  And this is exactly the regime when we expect a more phenomenological model (such as non-equilibrium thermodynamics) to provide some useful insight into the dynamics of information. In fact all this is not very different from how thermodynamics/hydrodynamics is used for modeling very large classical systems for which a precise microscopic description is either unknown or irrelevant. 

Before we proceed, let us also emphasize that in this paper we will make another explicit assumption, namely, that the physical space had emerged. As was already mentioned, it is not a priori clear how a low $D$-dimensional (e.g. $3$-dimensional) space can emerge from a vector in a very high dimensional Hilbert/statistical space. In Ref. \cite{graph_flow} we described one possibility (actually a large class of possibilities that we called the information graph flows) and to illustrate the procedure a mutual information was used as a measure of informational dependence, however other definitions of information matrix such as Fisher information matrix could have been used instead.\footnote{The key idea in Ref. \cite{graph_flow} was to approximate a state vector $|\psi\rangle$ or $\vec{p}$ by keeping track of only macroscopic parameters (such as information matrix) and neglecting microscopic parameters describing these vectors. If the dimensionality $2^N$ of the (Hilbert or statistical) space is very large, the number of parameters in the information matrix is still very large $\sim N^2$, but  one can employ the information graph flow procedure to reduce the number of parameters further. It was shown that for a suitable choice of the graph flow equations (with target dimensionality $D$) the number of (macroscopic) parameters in the information matrix (after the graph flow) can be as low as $\sim D N$. And then at the final stages of the graph flow it is more convenient to use an emergent spatial geometry and to define a local information tensor instead of the information matrix. At this stage the microscopic degrees of freedom (e.g. described by (qu)bits) already arranged themselves on a low $D$-dimensional lattice and can be thought to form macroscopic degrees of freedom described by positions of bits on the lattice. In the limit when the number of the qubits is large, the position variable can be assumed to take continuous values which is what we intend to do here.} In this paper we will consider a generalization of the Fisher information matrix (or the Fubini-Study metric) which better captures the quantum entanglements and also admits a covariant description. In fact, it might be useful (although not necessary) to consider the present discussion as the analysis of the final stages of the graph flow procedure when a $D$ dimensional space had already emerged, but a space-time covariance and an emergent dynamics of a metric is yet to be derived. However, instead of following the temporal dynamics of individual state vectors or individual information matrices as in Ref. \cite{graph_flow} we will work directly with, first, statistical ensembles as in Ref. \cite{Vanchurin} and, then, with thermodynamic/hydrodynamic variables.

The paper is organized as follows. In the next section we introduce and motivate the basic concepts of the (classical and quantum) information theories such as  Fisher information matrix and Fubini-Study metric. In Sec. \ref{sec:ensembles} we define statistical ensembles for a dual field which is identified with (or mapped to) either a complex wave-function or a square-roots of a probability distribution. In Sec. \ref{sec:computation} we construct a space-time covariant ensemble and define a fully covariant information-computation tensor which satisfies the conservation equations. In Sec. \ref{sec:thermodynamics} we develop (first equilibrium and then non-equilibrium) thermodynamic descriptions of the quantum/statistical systems. In Sec. \ref{sec:gravity} we show how the Einstein's dynamics of a metric might emergence due to a non-equilibrium production of entropy and argue that deviations from general relativity are expected far away from an equilibrium. In Sec. \ref{sec:discussion} we summaries the main results of the paper.

\section{Information Theory}\label{sec:information}

Consider a quantum  state $|\psi\rangle$ or a probability measure $\vec{p}$  and a preferred tensor product factorization of respectively Hilbert space $\cal{H}$,
\be
|x\rangle = \bigotimes_i |x_i\rangle
\ee
or statistical space $\cal{P}$,
\be
\hat{x} = \bigotimes_i \hat{x}_i
\ee
where $i \in \{1,...,D\}$ and $D$ is the dimensionality of a configuration (in the case of quantum states) or a sample (in the case of probability measures) space $\Omega$. (The dimensionality of either ${\cal H}$ or ${\cal P}$ will be denoted by $N$.) If the space $\Omega$ is compact and approximated by a lattice, then $N$ is finite and one can assume that $|x\rangle$ and $\hat{x}$ are normalized unit vectors with respect to the inner products on $\cal{H}$ and  $\cal{P}$ respectively. Then  we can define representations for $|\psi\rangle$ and $\vec{p}$ known respectively as a quantum wave-function
\be
\psi(x^1, ..., x^D)\equiv \langle x |\psi\rangle \label{eq:wave_function}
\ee
and a classical probability distribution
\be
p(x^1, ..., x^D) \equiv \vec{p} \cdot \hat{x}.
\ee
The only difference between the two objects is that the probability distribution $p(x^1, ..., x^D)$ is non-negative with normalization\footnote{Whenever the sample/configuration space $\Omega$ is approximated with a discrete lattice the integration should be interpreted as a summation over all lattice points.}
 \be
 \int d^D x \; p(x^1, ..., x^D)  =1
 \ee
 and the wave-function $\psi(x^1, ..., x^D)$ is a complex valued function with normalization
 \be
 \int d^D x \; \psi^*(x^1, ..., x^D)\psi(x^1, ..., x^D) =1.\label{eq:normalization}
 \ee

\subsection{Classical Information} 

A classical probability distribution $p(\vec{x})$ of independent random variables $x^1$, $x^2$, ... ,$x^D$ can always be factorized as 
\be
p(\vec{x}) = {p_1(x^1)} { p_2(x^2)} ... {p_D(x^D)}
\ee
where $\vec{x}=(x_1,...,x_D)$ is a vector in the sample space $\Omega$  (and not in $\cal P$). In the information theory we are usually dealing with dependent random variables and then it is important to have a covariant (in a sense of transforming covariantly between different tensor product factorizations of $\cal{H}$ and  $\cal{P}$) measure of the amount by which the probability distribution $p(\vec{x})$ does not factorize, i.e.
\be
p(\vec{x}) \neq {p_1(x^1)} { p_2(x^2)} ... {p_D(x^D)}.
\ee
By taking a logarithm of both sides we get
\be
\log(p(\vec{x})) \neq \log(p_1(x^1)) + \log(p_2(x^2)) + ... + \log(p_D(x^D)) \label{eq:logs}.
\ee
and by twice-differentiating with respect to $x_i$ and x$_j$  (such that $i\neq j $) we obtain a measure of statistical dependence between subsystems $i$ and $j$, i.e. 
\be
\Sigma_{ij}(\vec{x}) \equiv - \frac{\partial^2 }{\partial x^i \partial x^j }\log(p(\vec{x})) \neq - \frac{\partial^2 }{\partial x^i \partial x^j }\left ( \log(p_1(x^1)) + \log(p_2(x^2)) + ... + \log(p_D(x^D)) \right ) = 0.
\ee
Note that only off-diagonal elements of the Hessian matrix $\Sigma_{ij}(\vec{x})$ represent statistical dependence  between subsystems, and the meaning of the diagonal elements will become clear shortly. 

If the distribution  $p(\vec{x})$ is close to a Gaussian, then its logarithm can be expanded to the second order around global maxima at $\vec{y}$, i.e.
\be
\log(p(\vec{x})) = \log(p(\vec{y}))  - \frac{1}{2} (x^i - y^i) \Sigma_{ij}(\vec{y})  (x^j - y^j)+...,
\ee
and then 
\be
p(\vec{x}) \approx \sqrt{ \frac{|\Sigma|}{ \left ( 2 \pi \right )^{D} }}\;e^{- \frac{1}{2} (x^i - y^i) \Sigma_{ij}(\vec{y})  (x^i - y^i)}.
\ee
where $\Sigma$ is the determinant of $\Sigma_{ij}$.

As was already mentioned $\Sigma_{ij}$ describes (the amount of) informational dependence between $i$'s and $j$'s random variables (or subsystems), but we can do better than that. First note that the expansion of $\log(p)$ around an arbitrary local maxima at $y^{(J)}$ is still given by 
\be
\log(p(\vec{x})) \approx \log(p(\vec{y}^{(J)}))  - \frac{1}{2} \left (x^i - y_{(J)}^i\right ) \Sigma_{ij}(\vec{y}_{(J}))  \left (x^j - y_{(J)}^j \right )+...
\ee
and the entire probability distribution can be better approximated by a sum of Gaussians, i.e.
\be
p(\vec{x}) \propto \sum_J p(\vec{y}^{(J)}) \; e^{- \frac{1}{2} \left (x^i - y_{(J)}^i \right ) \Sigma_{ij}(\vec{y}_{(J})) \left (x^i - y_{(J)}^i \right )}.
\ee
 But now to obtain a single number (i.e. a measure) which represents informational dependence between $i$'s and $j$'s subsystems the Hessian matrix $ \Sigma_{ij}(\vec{y}_{(J})) $ must be summed (or integrated) over different $J$'s with perhaps different weights. One useful choice is to weight it by probability densities $p(\vec{y}_{(J}))$ and then in a continuum limit we get a good measure of informational dependence\footnote{Note that there might be other weight functions of the probability density $w(p)$ that could have been used for defining an information matrix, i.e. $A_{ij} = - \int d^D x\;\; w(p)\;\frac{\partial^2 }{\partial x^i \partial x^j }\log(p)$, but since the linear choice $w(p)= p$ seems to be the most natural (and also the simplest) we will use it throughout the paper.} 
\bea
A_{ij} &\equiv & \frac{1}{4} \int d^D x \;\; p(\vec{x}) \Sigma_{ij}(\vec{x})\notag \\
& =& - \frac{1}{4} \int d^D x\;\; p(\vec{x}) \;\frac{\partial^2 }{\partial x^i \partial x^j }\log(p(\vec{x})) 
\label{eq:Classical_information}
\eea 
where the factor of $1/4$ is introduced for future convenience. We will refer to $A_{ij}$ as a (classical) information matrix and will generalize it in the next subsection to better describe informational dependence in quantum systems. 

Note that if one considers a family of probability distributions $ p_\theta(\vec{x}; \theta^1, \theta^2, ... ) $ parametrized by some parameters $\theta^i$'s  then the following quantity is known as the Fisher information matrix 
\bea
F_{ij}  =  - \int d^D x\;\; p_\theta(\vec{x};  \theta^1, \theta^2, ... ) \;\frac{\partial^2 }{\partial \theta^i \partial \theta^j }\log(p_\theta(\vec{x}; \theta^1, \theta^2, ... )) .
\label{eq:Fisher_information}
\eea 
If we are to consider small shifts in the sample space coordinates, then this would generate a family of probability distributions parametrized by $D$ parameters $\theta^i$'s, i.e.
\be
p_\theta(\vec{x}; \theta^1, \theta^2, ... ) \equiv p(\vec{x} + \vec{\theta}) 
\ee
and then
\bea
F_{ij}  &=&  - \int d^D x\;\; p(\vec{x}+\vec{\theta}) \;\frac{\partial^2 }{\partial \theta^i \partial \theta^j }\log(p(\vec{x}+\vec{\theta}) ) \notag\\
&=&  - \int d^D x\;\; p(\vec{x}+\vec{\theta}) \;\frac{\partial^2 }{\partial x^i \partial x^j }\log(p(\vec{x}+\vec{\theta}) ).
\eea
This suggests that (up to a constant factor of $1/4$) our information matrix $A_{ij}$ is nothing but the Fisher information matrix for parameters $\theta^j$'s being shifts in the sample space coordinates $x^j$'s, i.e.
\be
A_{ij} = \frac{1}{4} \left [ F_{ij} (\vec{\theta})\right ]_{\theta^i=0}.
\ee
\subsection{Quantum Information}

Equation \eqref{eq:Classical_information} can be integrated by parts and (for vanishing or periodic boundary conditions) we obtain
\bea
A_{ij}   &=& - \frac{1}{4} \int d^D x\;\; p(\vec{x})\;\frac{\partial^2 }{\partial x^i \partial x^j }\log(p(\vec{x})) \notag \\
&  = &  \int d^D x\;\;  \frac{\partial \sqrt{p(\vec{x})}}{\partial x^i} \frac{\partial \sqrt{p(\vec{x})}}{\partial x^j} \notag\\
&= &  \int d^D x\;\;  \frac{\partial \psi(\vec{x})}{\partial x^i} \frac{\partial \psi(\vec{x})}{\partial x^j}
 \label{eq:Sigma2}
\eea
where $\psi$ is a real wave-function defined as 
\be
\psi(\vec{x}) \equiv \sqrt{p(\vec{x})}.
\ee 

For a complex wave-function  $\psi(\vec{x})$ (which represents quantum states \eqref{eq:wave_function}) one might also like to construct a matrix whose elements describe the informational dependence between subsystems.  By informational dependence (once again) we mean the amount by which the complex wave-function fails to factorize as in the Hartee-Fock approximation, i.e.
\be
\psi(\vec{x}) \neq \psi_1(x^1) \psi_2(x^2) ... \psi_D(x^D).
\ee
Then, by following derivations of the previous subsection,  we could have defined a new information matrix as
\bea
A_{ij} &=& - \frac{1}{4}\int d^D  x \psi(\vec{x})^* \psi(\vec{x}) \;\frac{\partial^2 }{\partial x^i \partial x^j }\log(\psi^*(\vec{x})\psi(\vec{x})) 
\notag \\
& =& \int d^D x \frac{\partial |\psi(\vec{x})| }{\partial x^i} \frac{|\partial \psi(\vec{x})|}{\partial x^j}.
\eea 
This, however, does not completely capture what we want, as there might still be non-factorization of phases which leads to quantum entanglements between subsystems. Perhaps a better object, which also carries information about phases, is a straightforward generalization of \eqref{eq:Sigma2}, i.e. 
\bea
A_{ij} \equiv  \int d^D x \frac{\partial \psi^*(\vec{x}) }{\partial x^i} \frac{\partial \psi(\vec{x})}{\partial x^j}.
\eea 

As was already mentioned, in the case of probability distributions our (classical) information matrix is related to the Fisher matrix and in the case of quantum states it is closely related to the Fubini-Study metric. Indeed, if we now consider small shifts in the configuration space coordinates $\vec{x}$, then this would generate a family of wave-functions parametrized by $D$ parameters $\theta^i$'s, i.e.
\be
\psi_\theta(\vec{x}; \vec{\theta}) \equiv \psi(\vec{x} + \vec{\theta}) 
\ee
In this case the Fubini-Study metric is given by
\bea
S_{ij}  &=&  \int d^D x \frac{\partial \psi^*(\vec{x}+ \vec{\theta}) }{\partial \theta^i} \frac{\partial \psi(\vec{x}+ \vec{\theta})}{\partial \theta^j}  - \left ( \int d^D x \frac{\partial \psi^*(\vec{x}+ \vec{\theta}) }{\partial \theta^i}{ \psi(\vec{x}+ \vec{\theta})}  \right )  \left ( \int d^D y \psi^*(\vec{y}+ \vec{\theta})   \frac{\partial \psi(\vec{y}+ \vec{\theta})}{\partial \theta^j}   \right ) \notag\\
&=& \int d^D x \frac{\partial \psi^*(\vec{x}+ \vec{\theta}) }{\partial x^i} \frac{\partial \psi(\vec{x}+ \vec{\theta})}{\partial x^j},
\eea
where the second term drops out for shifts which preserve normalization \eqref{eq:normalization} and thus
\be
A_{ij} = \left [ S_{ij} (\vec{\theta})\right ]_{\theta^i=0}.
\ee
Note that we could have considered other deformations of either probability distributions $\vec{p}(\vec{x})$ or wave-functions $\psi(\vec{x})$ that are not necessarily defined though shifts of either sample space or configuration space coordinates. In this more general case the Fisher information matrix and the Fubini-Study metric can be thought of as the covariant metric tensors on respectively a statistical manifold $\cal P$ and a Hilbert space $\cal H$. Instead of going this route and trying to construct covariant quantities in  $\cal P$ and $\cal H$, we will focus on the covariance in the sample/configuration space $\Omega$.

\section{Statistical Ensembles} \label{sec:ensembles}   

It is convenient to introduce a dual scalar field $\varphi(\vec{x})$ in the sample/configuration space $\Omega$ defined (up to a constant factor) as 
\be
\varphi(\vec{x})  \propto \begin{cases}
\sqrt{\hbar\;p(\vec{x})} \;\;\; &\text{for probability measures $\vec{p}$}\\
{\sqrt{\hbar}}\; \psi(\vec{x})\;\;\; &\text{for quantum states $|\psi\rangle $}
\end{cases}\label{eq:mapping} 
\ee
and then information matrix (from now on) is defined as
\bea
A_{ij} \equiv \frac{1}{\hbar} \int d^D x \frac{\partial \varphi^*(\vec{x}) }{\partial x^i} \frac{\partial \varphi(\vec{x})}{\partial x^j}.
\label{eq:Quantum_information}
\eea 
The parameter $\hbar$, which usually plays the role of Planck constant in quantum field theories and of temperature in statistical mechanics, need not to be either of the two in the context of the information theory considered so far. We will refer to the scalar field $\varphi$ as {\it infoton} due to its informational origin and to $A_{ij}$ as {\it information matrix} for both probability measures $\vec{p}$ and quantum states $|\psi\rangle $.  

The information matrix $A_{ij}$ is a global and non-covariant object in $\Omega$, but the integral in \eqref{eq:Quantum_information} suggests that one might be able to define a local and covariant quantity. For that we will consider probability distributions (known as second-order probabilities) over probabilities or over quantum states and so it might be useful to think of this as a ``second quantization''. More precisely, we will construct a statistical ensemble $P[\varphi]$ over the infoton field $\varphi$ which would define second-order probabilities through the equation
\be 
P[\vec{p}] = \int_{\varphi^2 \propto \hbar {p}} {\cal D} \varphi P[\varphi] \label{eq:real_field} 
\ee
or for quantum states through the equation
\be 
P[|\psi\rangle] = \int_{\varphi  \propto \sqrt{\hbar} \psi} {\cal D} \varphi {\cal D} \varphi^*  P[\varphi] \label{eq:complex_field} 
\ee
In both Eqs. \eqref{eq:real_field}  and \eqref{eq:complex_field} we only demand that the (real or complex) wave function is proportional to the (real or complex) infoton field $\varphi$, but an approximate normalization of $\varphi$ will be imposed later on and so it is useful to think of the infoton as either real or complex wave-function (up to a constant factor). 

In the language of statistical mechanics,  equations  \eqref{eq:real_field} and \eqref{eq:complex_field} describe what can be called a``micro-canonical'' ensemble with integration taken over all configurations of the infoton field $\varphi$ subject to ``microscopic'' constraint ${\varphi^2 \propto \hbar \vec{p}} $ or ${\varphi  \propto \sqrt{\hbar} \psi}$. In the following section we will define instead a ``canonical'' ensemble with ``macroscopic'' constrains set by expected values of the information matrix, $\bar{A}_{ij}$. Since the case of a complex infoton field is more general, for the most part we will assume that $\varphi$ is complex (i.e. corresponding to quantum states) and the case of a real $\varphi$ (i.e. corresponding to probability distributions) can be recovered by setting $\varphi=\varphi^*$.

\subsection{Canonical Ensemble}

Probability distributions over pure quantum states certainly contain more information about the system than density matrices (see, for example, Ref. \cite{Weinberg}), but that is not the main motivation to introduce them. What we want is the machinery to define  statistical ensembles over (microscopic) quantum states subject to (macroscopic) constraints such as the information matrix,  ${A}_{ij}$. More precisely, we wish to define a statistical ensemble over $\varphi$, and thus a probability distribution over $|\psi\rangle$'s, such that an expected information matrix is given by a Hermitian matrix
\be
\bar{A}_{ij} = \left \langle A_{ij}  \right \rangle.
\ee

In statistical mechanics ensembles are often defined using partition functions 
\be
{\cal Z} =  \int {\cal D} \varphi {\cal D} \varphi^* \;\;\exp \left ({-\frac{{\cal S}[\varphi]}{\hbar}} \right )\label{eq:real}
\ee
where the (Euclidean) action ${\cal S}$ is some functional of the infoton field $\varphi$. If a theory is local then the action ${\cal S}$ is given by an integral over a local function $\cal L$  of the fields and its derivatives, e.g.
\be
{\cal S}  =   \int d^D x\;\; {\cal L}\left (\varphi, \frac{\partial \varphi(\vec{x})}{\partial x^1}, ... ,\frac{\partial \varphi(\vec{x})}{\partial x^D} \right )
\ee
A simple example of a local theory is given by
\bea
{\cal S}  =   \int d^D x\;\; \left (g^{ij} \frac{\partial\varphi^*(\vec{x})}{\partial x^i} \frac{\partial\varphi(\vec{x})}{\partial x^j}  + \lambda  \varphi^*(\vec{x})\varphi(\vec{x}) \right ) \label{eq:Partition}
\eea
where the values of $g^{ij}$ do not (yet) depend on $\vec{x}$ and the ``mass-squared'' constant $\lambda$ must be chosen so that the infoton field (which is identified with wave-functions) is on average normalized.  The partition function defined by the action \eqref{eq:Partition} is a (functional) Gaussian and so it can be easily evaluated,
\bea
{\cal Z}  = \int {\cal D} \varphi{\cal D} \varphi^* \exp \left ( {- \frac{1}{\hbar} \int d^D x\;\; \varphi^* \left ( \lambda - g^{ij} \frac{\partial}{\partial x^i} \frac{\partial}{\partial x^j}  \right )\varphi } \right )= \det \left ( \frac{\lambda}{\hbar}- \frac{g^{ij}}{\hbar} \frac{\partial}{\partial x^i} \frac{\partial}{\partial x^j} \right )^{-1}.
\eea
The (functional) determinant is divergent and should be regularized, which is possible whenever $\lambda>0$ and $\det(g^{ij})>0$. The corresponding free energy is
\bea
{\cal F} &\equiv&  - \hbar \log({\cal Z}) \notag \\
&= & \hbar  \log\left (\det \left (\frac{\lambda}{\hbar}- \frac{g^{ij}}{\hbar} \frac{\partial}{\partial x^i} \frac{\partial}{\partial x^j}  \right ) \right ) \notag \\
&=&  \text{Tr} \left ( \hbar \log\left ( \frac{\lambda}{\hbar}- \frac{g^{ij}}{\hbar} \frac{\partial}{\partial x^i} \frac{\partial}{\partial x^j} \right ) \right )
\eea
and by differentiating it with respect to (Hermitian) $g^{ij}$ we get an expected information matrix
\be
\frac{1}{\hbar} \frac{\partial {\cal F}}{\partial g^{ij}}   = \left \langle \frac{1}{\hbar} \int d^D x\;\;  \frac{\partial}{\partial x^i}  \varphi^*\frac{\partial}{\partial x^j}\varphi  \right \rangle = \bar{A}_{ij}  \label{eq:cond1}
\ee
and an expected normalization 
\be
\frac{1}{\hbar}  \frac{\partial {\cal F}}{\partial \lambda} = \left \langle \frac{1}{\hbar} \int d^D x \;\; \varphi^*\varphi   \right \rangle =1. \label{eq:cond2}
\ee
Then the $N^2+1$ parameters in both $g^{ij}$ and $\lambda$ are to be chosen to satisfy $N^2+1$  equations \eqref{eq:cond1} and \eqref{eq:cond2}.
Thus one can view our ensemble as a ``canonical'' ensemble of (either quantum $|\psi\rangle$ or statistical $\vec{p}$) states subject to constraints imposed by expected information matrix $\bar{A}_{ij}$ and a normalization condition. 

\subsection{Information Tensor}

The next step is to restrict ourselves to only real (and symmetric)  $g_{ij}$, but to promote it to depend on local coordinates, which  amounts to improving our models of probability distributions from a single Gaussian to a (more general) sum of Gaussians. Then the information matrix $A_{ij}$ as well as its conjugate variables $g^{ij}$ should depend on  $(x^1, x^2, ..., x^D)$ and thus to play the role of a metric tensor. Moreover, we will add $\sqrt{|g|}$ to the volume integral and replace partial derivatives with covariant derivatives, i.e.
\be
{\cal S} = \int d^D x\;\sqrt{|g|}\; \left (g^{ij}(\vec{x}) \nabla_i \varphi^*(\vec{x}) \nabla_j  \varphi(\vec{x})  + \lambda \varphi^*(\vec{x})\varphi(\vec{x}) \right ).\label{eq:covariant_action}
\ee
There are two advantages of using this new partition function. The first one is that the ensemble is covariant and does not depend on the choice of coordinates. The second one is that it allows us to introduce a local (in the sample/configuration space $\Omega$) notion of informational dependence using variational derivatives of the free energy with respect to the metric $g^{ij}$. 

Indeed, if we define a covariant {\it information tensor} as
\be
{\cal A}_{ij}(\vec{x}) \equiv   \frac{1}{\hbar}  \nabla_{i} \varphi^*\nabla_{j}\varphi .\label{eq:information_tensor}
\ee
and a covariant  {\it probability scalar}
\be
{\cal N}(\vec{x}) \equiv\frac{1}{\hbar} \varphi^*(\vec{x})\varphi(\vec{x}),
\ee
then these two quantities can be used to express the stress tensor for the infoton field as
\bea
T_{ij}(\vec{y}) =\nabla_{(i} \varphi^*\nabla_{j)}\varphi  + g_{ij} \left (g^{kl} \nabla_k \varphi^*  \nabla_l \varphi  + \lambda  \varphi^*\varphi \right ) \\
= 2 \hbar {\cal A}_{(ij)} + \hbar g_{ij} \left (g^{kl} {\cal A}_{kl} + \lambda  {\cal N}   \right )
  \label{eq:Tij}
\eea
where symmetrization (and for later use anti-symmetrization) of indices is defined as
\bea
{\cal A}_{[\mu\nu]} &\equiv& \frac{1}{2}\left ( {\cal A}_{\mu\nu} -{\cal A}_{\mu\nu}   \right ) \notag\\
{\cal A}_{(\mu\nu)}&\equiv& \frac{1}{2}\left ( {\cal A}_{\mu\nu} +{\cal A}_{\mu\nu}   \right ) .\label{eq:symmetrization}
\eea
Then for a given (ensemble averaged) information tensor $\bar{\cal A}_{(ij)}$ the parameters $g_{ij}(\vec{x})$ and $\lambda$ are to be chosen such that 
\be
\int d^D x \;\sqrt{|g(\vec{x})|} \;\;\left \langle {\cal N}(\vec{x}) \right \rangle  =  \int d^D x  \;\;
\frac{1}{\hbar}  \frac{\delta {\cal F}[g,\lambda]}{\delta \lambda}  =- \int d^D x   \frac{\delta \log \left ({\cal Z}[g,\lambda]\right )}{\delta \lambda}   =1
\ee
and
\be
 \left \langle T_{ij }(\vec{y}) \right \rangle =  \frac{2}{\sqrt{|g(\vec{y})|} } \frac{\delta {\cal F}[g,\lambda]}{\delta g^{ij} (\vec{y}) } = 2 \hbar \bar{\cal A}_{(ij)} + \hbar g_{ij} \left (g^{kl} \bar{\cal A}_{kl} + \lambda \left \langle {\cal N}  \right \rangle \right ).
\ee
One can also obtain the expected equation of motion for the infoton field by extremizing the action, i.e.
\be
\frac{\delta {\cal S}[\varphi]}{\delta \varphi^*}  = \left ( \lambda - g^{ij} \nabla_i \nabla_j\right ) \varphi=0.\label{eq:classical_eom}
\ee

\section{Computation Theory}\label{sec:computation}

This completes our discussion of statistical ensembles over either quantum states or probability distributions in the context of either quantum or classical information theory. The next step is to generalize the construction to ensembles over trajectories in either Hilbert or statistical spaces. As before, the main idea will be to keep track of only a small number of macroscopic parameters (such as expected information tensor) at all times, but in addition we will also restrict the dynamics to only {\it local computations} that we define next.

\subsection{Local Computations}\label{sec:local_computation}

Consider a quantum sate $|\psi\rangle$ of a computational system of $D$ qutrits or more generally qudits. The Hilbert space of a single qudit has and a fixed number $K$ of dimensions that in addition we also assume to have a cyclic ordering. For $K=2$ a qudit is just a qubit, for $K=3$ a qudit is a qutrit and the (cyclic) ordering of dimensions is not required, but for $K>3$ it is essential for what we are going to do next. If one is not yet comfortable with the concept of qudits, it is safe to assume (for starters) that $K=3$ and the necessity of ordering of dimensions and of introduction of qudits will become clear shortly.  Then a sample/configuration space $\Omega$ for the system can be thought of as a  $D$ dimensional lattice with $K^D$ lattice points (e.g. $3^D$ for qutrits) and periodic boundary conditions, i.e. a $D$ dimensional torus (see Fig. \ref{fig:torus}.)
\begin{figure}[]
\begin{center}
\includegraphics[width=0.6\textwidth]{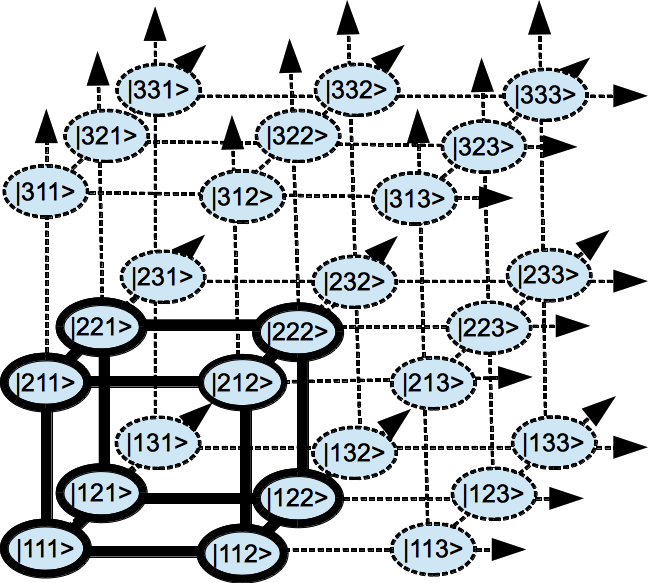}
\caption{Physical space for a dual field theory of a system of three q-trits (i.e.  three qudits with $K=3$.) Lattice points are marked with respective computational basis vectors and a single copy of a lattice hypercube is highlighted with solid bold lines.  \label{fig:torus}}
\end{center}
\end{figure}

Indeed, the computational basis are given by
\be
|n\rangle = \bigotimes_i | n_i\rangle
\ee
where $n_i \in \{1, 2, ..., K\}$ and thus $n$ is an integer in base $K$, i.e. $n \in\{1, 2, ... , K\}^D$ and then the wave-function is defined as
\be
\psi(n) \equiv \langle n | \psi \rangle. 
\ee
In our dual field theory picture the system is described with the infoton field
\be
\varphi(n) \propto \sqrt{\hbar} \psi(n)
\ee
where the integer $n$ (in base $K$) specifies lattice coordinates. Note that for qutrits (or $K=3$) the labeling of digits with $1$, $2$ and $3$  has no significance,  and the (dual) description should be invariant under arbitrary permutations of these labels. To accommodate this symmetry (on the dual field theory side) we assume that the $D$ dimensional lattice is periodic, i.e. $D$-torus. Then, by construction, $n^i=1$ is related to $n^i=2$ in exactly the same way as $n^i=2$ related to $n^i=3$ or as $n^i=3$ related to $n^i=1$. For $K>3$ the (cyclic) ordering of dimensions in our definition of qudits is the necessary additional ingredient that allows us to use an integer $n$ in base $K$ as a coordinate on a $D$-dimensional periodic lattice and thus allows us to define a physical space for the dual field theory. If the ordering was not available and $K>3$ we might still be able to define the dual physical space, but its dimensionality would have to be larger than $D$.

In what follows we will look for a computational dynamics of wave-functions which have a dual description in terms of generally covariant theories for the infoton field. As one might expect such dynamics of wave-functions will be even more restrictive than allowing only a few (one or two) qudit gates. Indeed even applications of a single qudit gate would generically effect all lattice points. However, if we are restricted to only local changes of the infoton field, then we are to constrain ourselves to a small subset of quantum gates on qudits. We will call these gates {\it local} gates, but it is still expected that these gates will form a universal set of quantum gates. (This will be easy to see on the dual field theory side where solutions for the infoton field must exist for an arbitrary inital and final configurations of the field.) In order to understand what these gates are, let us zoom-in on a single $D$-dimensional lattice hypercube of our $D$-dimensional lattice (see Fig. \ref{fig:torus}). Since the lattice is invariant under shifts by any of the lattice vectors, we can (without loss of generality) take the hypercube to have vertices with coordinates $n \in \{1,2\}^D$ (highlighted with solid bold lines on Fig. \ref{fig:torus}). This is just a hypercube of the very same type that was used in Ref. \cite{Vanchurin} with the difference that the boundary conditions are not periodic. In other words, if you are to move along $i$'s dimension from $n^i=1$ to $n^i=2$ and then continue the motion, then you will not come back to $n^i=1$ until after you pass through $n^i= 3, ...K$. But now if we are interested in all possible transformations of the infoton on this (local) hypercube, this is exactly the realm of quantum computing for a system of $D$ qubits. It is well known that to describe all such transformation we can restrict ourselves to only one and two qubit gates (which do from a universal set of gates), and those are exactly the {\it local computations} we are interested in. 

In summary, the local computations in each of $K^D$ hypercubes are to be described by (local one or two) qubit gates but these computations are not isolated from each other and not only run in parallel, but can communicate with each other by exchanging information and also results of computations. For a reader familiar with parallel computing it might be more insightful to think of these hypercubes as  $K^D$ quantum computers that are arranged on a lattice $D$-dimensional lattice with periodic boundary conditions that run in parallel and share memory with their neighbors.\footnote{Note that we do not know how or if this architecture can be realized in practice, but the picture of the parallel local computations is useful for deriving a covariant description of information and computations.} For example a quantum computer described by a $D$-dimensional hypercube $n \in \{1,2\}^D$ and a neighboring quantum computer described by a  $D$-dimensional  hypercube $n \in \{1,2\}^{D-1}\times \{2,3\}$ share memory (or Hilbert subspace) described by a  $D-1$-dimensional  hypercube $n \in \{1 ,2\}^{D-1} \times \{ 2\}$. Clearly, in such a network local information (or results of local computations) cannot propagate instantaneously across the entire torus which is exactly what we need for a Lorentz invariance to emerge on a dual field theory side. In the following subsection we will construct a simple space-time ensemble described by a local Lagrangian of the infoton field, but it is important to keep in mind that the dual theory can be (and perhaps should be) further complicated by additional gauge fields as in \cite{Vanchurin} or by an (emergent) metric field as in Sec. \ref{sec:gravity}.

\subsection{Space-time Ensembles} 

Consider a probability distribution $p(\vec{x})$ or a wave-function $\psi(\vec{x})$ which evolves in time, i.e. it depends on $D$ spatial coordinates $x^i$ and one time coordinate $x^0$. As before we are interested to see how these functions fail to factor into products, i.e.
\bea
p(\vec{x}) &\neq& {p_1(x^0, x^1)} { p_2(x^0, x^2)} ... {p_D(x^0, x^D)}\\
\psi(\vec{x}) &\neq& {\psi_1(x^0, x^1)} { \psi_2(x^0, x^2)} ... {\psi_D(x^0, x^D)},
\eea 
where by $\vec{x}$ we now denote a $D+1$ dimensional vector in sample/configuration space-time, i.e. $(x^0, x^1,...,x^D)$. What is, however, different is that  an approximate normalization of the infoton field must be imposed independently at all times $x^0$. This can be accomplished with the following (non-covariant) action 
\be
\tilde{\cal S}  =   \int d^{D+1} x\sqrt{|g|} \left (g^{ij}(\vec{x}) \frac{\partial \varphi^*(\vec{x})}{\partial x^i} \frac{\partial \varphi(\vec{x})}{\partial x^j}    +  \lambda(x^0) \varphi^*(\vec{x})\varphi(\vec{x}) \right ) 
 \label{eq:action3}
\ee
where the tildes are introduced to indicate that we are now working with space-time quantities. As before the first term defines a statistical ensemble for desired information tensor, but the second term is placed to ensure that probabilities are conserved at all times, i.e.
\be
\frac{1}{\hbar} \frac{\delta \tilde{\cal F}}{\delta \lambda(y^0)} = \left \langle \frac{1}{\hbar} \int d^D x \;\; \varphi^*\varphi  \right \rangle =1. \label{eq:lambda}
\ee
Upon variation of the action with respect to the infoton field we get the expected equation of motion
\bea
 \left ( \lambda(x^0) - g^{ij}(\vec{x}) \nabla_i \nabla_j  \right ) \varphi(\vec{x})=0,
\eea
where by $\nabla_i$ we denote (spatially) covariant derivatives. This is nothing but a time-independent Schrodinger equation and so the evolution is entirely due to the time dependence of $\lambda(x^0)$, i.e. external source. 

After integration by parts the action of \eqref{eq:action3} can be rewritten as 
\be
\tilde{\cal S}  =  \int d^{D+1} x\sqrt{|g|} \left (g^{ij}(\vec{x}) \frac{\partial \varphi^*(\vec{x})}{\partial x^i} \frac{\partial \varphi(\vec{x})}{\partial x^j}    -  2 \lambda^{(1)}(x^0) \Re \left ( \varphi^*(\vec{x}) \frac{\partial }{\partial x^0}  \varphi(\vec{x}) \right ) \right ) \label{eq:action4}
\ee
where 
\be
\lambda^{(1)}(x^0) \equiv \int_{0}^{x^0} d\tau\lambda(\tau).
\ee
This new form suggests another statistical ensemble that we can construct by simply replacing $\Re$ (or real) with $\Im$ (or imaginary) in the action, i.e.
\bea
\tilde{\cal S}  & = &    \int d^{D+1} x\sqrt{|g|} \left (g^{ij}(\vec{x}) \frac{\partial \varphi^*(\vec{x})}{\partial x^i} \frac{\partial \varphi(\vec{x})}{\partial x^j}    -  2 \lambda^{(1)}(x^0) \Im \left ( \varphi^*(\vec{x}) \frac{\partial }{\partial x^0}  \varphi(\vec{x}) \right ) \right ) \label{eq:action5} \\ 
 & = &   \int d^{D+1} x\sqrt{|g|} \left (g^{ij}(\vec{x}) \frac{\partial \varphi^*(\vec{x})}{\partial x^i} \frac{\partial \varphi(\vec{x})}{\partial x^j}    + i \lambda^{(1)}(x^0) \left ( \varphi^*(\vec{x}) \frac{\partial }{\partial x^0}  \varphi(\vec{x})  - \varphi(\vec{x}) \frac{\partial }{\partial x^0}  \varphi^*(\vec{x}) \right ) \right )  \notag
\eea
Upon variation the expected equations of motion are now given by 
\bea
 \left (i \lambda^{(1)}(x^0)\frac{\partial }{\partial x^0} - g^{ij}(\vec{x}) \nabla_i \nabla_j  \right ) \varphi(\vec{x}) =0.
\eea
which is a time-dependent Schrodinger equation. Now the dynamics is not only due to time-dependence of  $\lambda^{(1)}$, but also due to internal dynamics of the infoton field $\varphi$.

If we try to promote the action \eqref{eq:action5} to a covariant form we encounter a problem which was first noticed by Dirac: the action has terms with only one time derivative, but two spatial derivative. We are not going to ``fix'' it using Dirac  spinors. Instead, we will first define a different space-time ensemble for the infoton field and then we will argue that local computations (defined above \ref{sec:local_computation}) would give rise to such an ensemble.

Up to a boundary term the action \eqref{eq:action5}  can be rewritten as
\bea
\tilde{\cal S} & = & \int d^{D+1} x\sqrt{|g|} \left ( g^{ij}(\vec{x}) \frac{\partial \varphi^*(\vec{x})}{\partial x^i} \frac{\partial \varphi(\vec{x})}{\partial x^j}    - i \lambda^{(2)}(x^0) \left ( \varphi^*(\vec{x}) \frac{\partial^2 }{\partial (x^0)^2}  \varphi(\vec{x})  - \varphi(\vec{x}) \frac{\partial^2 }{\partial (x^0)^2}  \varphi^*(\vec{x}) \right ) \right )\notag\\
& = &\int d^{D+1} x\sqrt{|g|} \left (  g^{ij}(\vec{x}) \frac{\partial \varphi^*(\vec{x})}{\partial x^i} \frac{\partial \varphi(\vec{x})}{\partial x^j}    +2 \lambda^{(2)}(x^0) \Im \left ( \varphi^*(\vec{x}) \frac{\partial^2 }{\partial (x^0)^2}  \varphi(\vec{x}) \right ) \right ),
\eea
where
\be
\lambda^{(2)}(x^0) \equiv \int_{0}^{x^0}  d\tau  \lambda^{(1)}(\tau)= \int_{0}^{x^0}  d\tau \int_{0}^{\tau} d\sigma  \lambda(\sigma).
\ee
Then by (once again) replacing $\Im$ with $\Re$ we define a new ensemble described by action
\bea
\tilde{\cal S}  & = &  \int d^{D+1} x\sqrt{|g|} \left (g^{ij}(\vec{x})  \frac{\partial \varphi^*(\vec{x})}{\partial x^i} \frac{\partial \varphi(\vec{x})}{\partial x^j}    +2 \lambda^{(2)}(x^0) \Re \left ( \varphi^*(\vec{x}) \frac{\partial^2 }{\partial (x^0)^2}  \varphi(\vec{x}) \right )\right ) \label{eq:action6} \\ 
 & = &  \int d^{D+1} x\sqrt{|g|} \left (g^{ij}(\vec{x})  \frac{\partial \varphi^*(\vec{x})}{\partial x^i} \frac{\partial \varphi(\vec{x})}{\partial x^j}    +\lambda^{(2)}(x^0) \left ( \varphi^*(\vec{x}) \frac{\partial^2 }{\partial (x^0)^2} \varphi(\vec{x}) +\varphi(\vec{x}) \frac{\partial^2 }{\partial (x^0)^2}  \varphi^*(\vec{x}) \right )
 \right ),  \notag
\eea
which has the desired property of having the same number of spatial and time derivatives and so can be covariantized,
\bea
\tilde{\cal S}  & = &  \int d^{N+1} x\;\sqrt{|g|} \;\; g^{\mu\nu} \nabla_\nu \varphi^*(\vec{x})  \nabla_\mu \;\varphi(\vec{x}) 
 \label{eq:fullly_covariant_action} 
\eea
where $g^{00} =- 2 \lambda^{(2)}$, $g^{i0}$ and $g^{ij}$ are all promoted to be functions inn sample space and also time. If $\lambda$ is positive, both $ \lambda^{(1)}$ and  $\lambda^{(2)}$ are positive and so the metric has Lorentzian signature and therefore the corresponding partition function must be complex, i.e.
\be
{\cal Z} =  \int {\cal D} \varphi {\cal D} \varphi^* \;\;\exp \left ({i\frac{\tilde{\cal S}[\varphi]}{\hbar}} \right )\label{eq:complexl}.
\ee

An important observation is that this partition function describes the simplest example of a dual field theory of quantum computation discussed in Ref. \cite{Vanchurin}. \footnote{Note that in Ref. \cite{Vanchurin} the dual theory had to be generalized further to include additional fields (e.g. an Abelian field) in order to cure the so-called $Z$-problem. We will consider such theories in Sec \ref{sec:gravity}, but the additional (emergent) field will be the metric tensor $\mg_{\mu\nu}$.} The fact that \eqref{eq:fullly_covariant_action} admits a space and also time covariance can be understood by recalling a condition/restriction that was imposed on the dynamics of wave-functions on a quantum computation side in Ref. \cite{Vanchurin} and discussed above in subsection \ref{sec:local_computation}. The condition was a suppression of the dynamics of wave-function which involves non $k$-local (i.e. acting non-trivially on many qubits) interactions/computations. The main difference, however, is that in the case of Ref. \cite{Vanchurin} the physical space consisted of a single Hamming cube, but in a more general case, considered in subsection \ref{sec:local_computation}, it can be arbitrary large (as on Fig. \ref{fig:torus}). With this respect the theory described by action \eqref{eq:fullly_covariant_action} should be viewed as a dual field theory of computation where the fundamental units of information need not be qubits, but can be qutrits and more generally qudits.
 
Upon variation of the action \eqref{eq:fullly_covariant_action} (with possibly potential term) with respect to the infoton field $\varphi$ we obtain a fully covariant equation, 
\be
\nabla_\mu \nabla^\mu \varphi = \frac{V(|\varphi|)}{\partial \varphi^*}. \label{eq:equation_of_motion}
\ee
As was already known to Dirac, not all of the solutions of this equation can be interpreted as wave-functions (or square-roots of probabilities), which is not a big problem.  In our analysis we view $\varphi$ not as a wave-function, but as a dual field whose role is to describe an ensemble from which an approximate dynamics of wave-functions is obtained. On the other hand, it is well known that in a non-relativistic limit the solutions of $\varphi$ would be described by a Schrodinger equation and so can be interpreted as wave-functions/square-roots of probabilities (see Ref. \cite{Vanchurin} for discussion of this point in context of dual  theories of quantum computation). 

Also note that the equation of motion \eqref{eq:equation_of_motion}} is of the second order and so a unique solution must exists for arbitrary configurations of initial and final states of the infoton field, $\varphi$. This is a dual field theory statement. On the quantum computation side, this result implies that there must exist a circuit composed of only local gates which are capable of performing arbitrary transformations, i.e. transformations between arbitrary quantum states, $|\psi\rangle$. Of course, this is not a prove, but only an evidence, suggesting that local gates do form a universal set of gates as was already mentioned.

\subsection{Information-Computation Tensor}

A space-time covariant action for the infoton field  \eqref{eq:fullly_covariant_action} provides us with a dual description of not only information, but also computation. It also suggests a covariant generalization of the information tensor, i.e.
\be
{\cal A}_{\mu\nu} \equiv  \frac{1}{\hbar}  \nabla_\nu\varphi^*\nabla_\mu\varphi.
\ee
We had already seen that the space-space components, i.e. ${\cal A}_{ij}$, provide a good measure of informational dependence between (local) quantum or statistical subsystems, but now there are also time-time,  i.e. ${\cal A}_{00}$, and space-time components, i.e.  ${\cal A}_{0i}$, that should also have a dual interpretation. In particular, the time-time component  measures temporal changes of the infoton field that can only take place due to computations on the information side, but these computations cannot occur unless the wave-function (represented by the infoton field) has spatial gradients. This suggests that the space-time component ${\cal A}_{0i}$ (which is a product of a spatial derivative and a temporal derivative of the infoton field) measures the amount that a given random variable $i$ is contributing to the computation in ${\cal A}_{00}$. Thus it is useful to think of ${\cal A}_{00}$ as a {\it density of local computations} and of ${\cal A}_{0i}$ as a {\it flux of local computations} which together with information tensor ${\cal A}_{ij}$ form a generally covariant tensor ${\cal A}_{\mu\nu}$ that we shall call the {\it information-computation} tensor. In subsection \ref{sec:local_computation} we already described a flow of information and also computations by constructing a network of quantum computers arranged on a $D$ dimensional lattice running in parallel and sharing memory resources between neighboring computers.\footnote{Note that we do not claim that such a network can be realized in practice, but it does allow us to introduce the notion of local information and computations.} Now we see that on a dual field theory side this is encoded in the information-computation tensor, ${\cal A}_{\mu\nu}$ .

As such, the information-computation tensor is not conserved, but it is related to the energy momentum tensor of the infoton field 
\be
T_{\mu\nu} = - 2 \hbar {\cal A}_{(\mu\nu)} + \hbar g_{\mu\nu} \left (g^{\alpha\beta} {\cal A}_{\alpha\beta}  \right )
\ee
which is conserved, i.e.
\be
\nabla^\mu T_{\mu\nu} =0.
\ee
This implies that the information-computation tensor should satisfy the following (conservation) equation
\be
\nabla^\nu  \left ( {\cal A}_{(\mu\nu)} - \frac{1}{2} {\cal A}  g_{\mu\nu} \right )  = 0
\ee
or 
\be
\nabla^\nu {\cal A}_{(\mu\nu)} = \frac{1}{2} \nabla_\mu {\cal A}.
\ee
where ${\cal A} \equiv g^{\alpha\beta} {\cal A}_{\alpha\beta}$ is the trace.

Now that we obtained a fully generally covariant description of information (and also computation) in terms of dual field theories (what can be called the second quantization of information), we could proceed to the third, fourth and so on quantizations. In the language of probability theory, we can look at third-order, forth-order and so on probabilities. We are not going to take this route here and instead we will switch to discussing an emergent dynamics of the metric tensor, $g_{\mu\nu}$. The reason why we expect $g_{\mu\nu}$ to be dynamic (or more precisely thermodynamic) should become clear shortly.

\section{Thermodynamics}  \label{sec:thermodynamics}

Let us go back to the action described by  \eqref{eq:covariant_action} with only spatial covariance. Since the corresponding free energy only depends on $g^{ij}(\vec{x})$ and $\lambda(\vec{x})$  it can be expanded as 
\bea
{\cal F}&\approx &\int d^D x \sqrt{|g|} \left (g^{ij} \left \langle {\cal A}_{ij} \right \rangle + \lambda \left \langle {\cal N} \right \rangle  \right )  - S.
\label{eq:Free_energy}
\eea
where $\hbar=1$ and $S$ is some constant. This free energy provides an approximate description of our ``canonical'' ensemble of states, but if we turn on (a unitary, but perhaps unknown) dynamics for wave-functions then the dual infoton field should also evolve accordingly. As a result the ensemble and expectation values $\left \langle {\cal A}_{ij} \right \rangle$ and  $\left \langle {\cal N} \right \rangle $ will change in some way. However, if we still want to keep the form of the ensemble to remain canonical, then, what must change are the macroscopic parameters $g_{ij}$ and $\lambda$ in  \eqref{eq:covariant_action}.  And if so, would it be possible to describe the emergent dynamics of $g_{ij}$ using dynamical equations, e.g. Einstein equations? In Sec. \ref{sec:gravity} we will attempt to answer this question, but instead of considering a specific unitary evolution for the wave-function (and the corresponding approximate evolution for the infoton field) we will consider random local computations (defined in subsection \ref{sec:local_computation}). Then one should be able to apply the ideas of hydrodynamics to obtain an effective equation for the metric that would be treated as a thermodynamic variable.

\subsection{Local Equilibrium}

To emphasize that we are now dealing with (local) thermodynamic variables we define the following quantities:
\bea
\text{information tensor} \;\;\; \ma_{ij} & \equiv & \left \langle {\cal A}_{ij} \right \rangle \\
\text{metric tensor} \;\;\; \mg^{ij} & \equiv & g^{ij} \\
\text{particle number scalar} \;\;\; \mn & \equiv & \left  \langle {\cal N} \right \rangle \\
\text{chemical potential scalar} \;\;\; \mm & \equiv & \lambda  \\
\text{entropy scalar} \;\;\; \ms & \equiv & \frac{S}{\int d^D x \sqrt{|g|}} \\
\text{free energy scalar} \;\;\; \mf & \equiv &  \left (   \mg^{ij} \ma_{ij} + \mm \mn \right ) - \ms  \label{eq:Euler_equation} 
\eea
and then the total free energy can be expressed as
\be
{\cal F} = \int d^D x \sqrt{|g|} \; \mf.
\ee
Equation \eqref{eq:Euler_equation} is the Euler equation (for thermodynamics) and then by combining it with the first law of thermodynamics
\bea
0 &=&    \mm  \td \mn+\mg^{ij}  \td \ma_{ij} - \td \ms  
\label{eq:first_law} 
\eea
we get the Gibbs-Duhem Equation 
\be
 \td \mf = \mn  \td \mm+\ma_{ij}  \td \mg^{ij}  \label{eq:Gibbs-Duhem}.
\ee
Note that the quantity in parenthesis in \eqref{eq:Euler_equation}  is usually interpreted as internal energy scalar,
\be
\me \equiv  \mm \mn+\mg^{ij} \ma_{ij}.
\ee
but in our case we not only know that it is a scalar as a whole, but we also know how it is  partitioned between different components, i.e. $\mm \mn$, $\mg^{11} \ma_{11}$, $\mg^{12} \ma_{12}$, etc. To summarize, the three thermodynamics equations (defined locally) are \bea
\text{Euler Equation:}&\;\;\;\;\;\; &\ms+\mf  =    \mm \mn +\mg^{ij} \ma_{ij}.\label{eq:covariant_Euler} \\
\text{First Law of Thermodynamics:} &\;\;\;\;\;\;& \td \ms  =   \mm \td \mn+\mg^{ij} \td\ma_{ij}  \label{eq:covariant_first_law}\\
\text{Gibbs-Duhem Equation:}& \;\;\;\;\;\;&   \td \mf = \mn  \td \mm+\ma_{ij}  \td \mg^{ij}  \label{eq:gibbs_duhem} 
\eea
Evidently, the free energy scalar $\mf$ is an extensive quantity, but the total free energy $\cal F$ is neither extensive nor intensive as it involves a larger number of (local) thermodynamic system which are not in a (global) equilibrium with each other. Of course, the expectation is that when the system is equilibrated the total free energy would become an extensive quantity.

We can use the definition of free energy to express a partition of unity 
\be
1  = \int {\cal D} \varphi{\cal D} \varphi^* \exp  \left ( -{\cal S} [\mg, \mm, \varphi] \right )
\ee
where
\be
{\cal S} [\mg, \mm, \varphi] \equiv \int d^D x\;\sqrt{|\mg|} \; \left ({\cal L}(\varphi, \mg, \mm) -  \mf(\mg, \mm)  \right ). \label{eq:entropy}
\ee
is the entropy functional which (in the equilibrium) does not depend on $\mm$ nor $\mg$, but is a function of $\varphi$.  However, according to \eqref{eq:covariant_Euler}, on average 
\be
\left \langle {\cal S} [\mg, \mm, \varphi] \right \rangle =   \int d^D x\;\sqrt{|\mg|} \;\left(  \mm \mn +\mg^{ij} \ma_{ij} -\mf  \right ) =  \int d^D x\;\sqrt{|\mg|} \;  \ms
\ee
which is independent on $\varphi$. Moreover, the dependence also vanishes near field configurations which satisfy classical equations of motion, i.e
\be
\frac{\delta {\cal S} [\mg, \mm, \left \langle \varphi\right \rangle ] }{\delta \left \langle \varphi\right \rangle} = 0.
\ee

\subsection{Non-equilibrium Thermodynamics}
 
The next step if to understand non-equilibrium effects \cite{Landau}. In that case instead of dealing directly with entropy \eqref{eq:entropy} we are interested in entropy production, i.e.
\be
\frac{\td}{\td t} {\cal S} [\mg, \mm, \varphi] =  \int d^D x\;\sqrt{|\mg|} \; \frac{\td}{\td t} \left ({\cal L}(\varphi, \mg, \mm) -  \mf(\mg, \mm)  \right ). \label{eq:entropy_production}
\ee
or in a more (but not completely) covariant from
\bea
{\cal S}  [\mg, \mm, \varphi]  &=& \int d^{D} x dt \;\sqrt{|\mg|} \; \left ( \mv^\alpha \nabla_\alpha \left ({\cal L}(\varphi, \mg, \mm)  - \mf(\mg, \mm)\right )  + \Lambda \right ) 
\label{eq:action2}
\eea
where $\mv_\nu$ defines a local rest frame of the fluid and $\Lambda$ is some integration constant. The main idea behind non-equilibrium thermodynamics is that the dynamics of systems near equilibrium is such that the entropy production is minimized.  Apart from the integration constant  $\Lambda$  there are two terms in the entropy functional: the first one represents the dynamics of the infoton field, i.e. 
\be
 \int d^{D+1} x \;\sqrt{|\mg|}  \mv^\alpha \nabla_\alpha {\cal L}(\varphi, \mg, \mm) 
 \ee
 and the other one is supposed to describe the dynamics of the metric, i.e. 
 \be
 - \int d^{D+1} x \;\sqrt{|\mg|}  \mv^\alpha \nabla_\alpha \mf(\mg, \mm). 
 \ee

In the previous section we argued that a full  space-time covariance on a field theory side arose from restricting to only local computations on the information theory side. In that case the corresponding (Euclidean space-time) free energy can also be used to partition unity, i.e.
\be
1  = \int {\cal D} \varphi{\cal D} \varphi^* \exp  \left ( -\tilde{\cal S} [\tilde{\mg}, \varphi] \right )
\ee
where
\be
\tilde{\cal S} [\tilde{\mg}, \varphi] \equiv \int d^{D+1} x\;\sqrt{|\tilde{\mg}|} \; \left (\tilde{\cal L}(\varphi, \tilde{\mg}) -  \tilde{\mf}(\tilde{\mg})  \right ) \label{eq:action}
\ee
and all the tildes are to remind us that these are space-time quantities. In particular $\tilde{\mg}$ is a space-time metric which (as was argued in the previous section) combines both the spatial metric $\mg$ and (integrals of) $\mm$. Thus, if we are to consider a random dynamics of quantum $|\psi\rangle$ (or statistical $\vec{p}$) states constrained to only local computations, then the entropy/action functional must be covariantized, i.e.
\be
{\cal S}  [\mg, \mm, \varphi] \rightarrow \tilde{\cal S} [\tilde{\mg}, \varphi].
\ee
This suggests that we should make the following identifications
\bea
\mv^\alpha \nabla_\alpha {\cal L}(\varphi, \mg, \mm) &\;\;\;\;\rightarrow \;\;\;\;\;& \tilde{\cal L}(\varphi, \tilde{\mg}) \\
 \mv^\alpha \nabla_\alpha \mf(\mg, \mm) + \Lambda & \rightarrow & \tilde{\cal \mf}(\tilde{\mg})
 \eea
and so the only local term which so far does not have a space-time covariant description is the entropy production term
\be
\mv^\alpha \nabla_\alpha \mf(\mg, \mm) \;\;\;\;\rightarrow \;\;\;\; \frac{1}{2\kappa}{\mR}(\tilde{\mg}) \equiv \tilde{\mf}(\tilde{\mg}) - \Lambda.
\ee
In the equilibrium $\td \mf = \mv^\alpha \nabla_\alpha \mf$ vanishes and so $\frac{1}{2\kappa}{\mR}(\tilde{\mg}) $ is due entirely to non-equilibrium processes. Note that $1/2$ and $\kappa$ are introduced for future convenience and at this point have no physical significance. 

\section{Emergent Gravity}\label{sec:gravity} 

So far we have only concluded that a fully space-time covariant entropy/action functional is expected to have the following form
\be
{\cal S} [{\mg}, \varphi] \equiv \int d^{D+1} x\;\sqrt{|{\mg}|} \; \left ({\cal L}(\varphi, {\mg}) - \frac{1}{2\kappa}{\mR}({\mg}) +\Lambda \right )\label{eq:full_action}
\ee
where we dropped all of the tildes, but it is (from now on) assumed that all of the quantities are defined in space-time. The term $\frac{1}{2\kappa}{\mR}({\mg})$ is responsible for non-equilibrium entropy production, but it is now convenient to think of this terms as (a yet to be derived) action for the metric degrees of freedom, although as we have seen the metric ${\mg}_{\mu\nu}$ is not a fundamental field, but a thermodynamic variable.

To describe a phenomenological procedure that we shall apply, let us first consider (the four-vectors of) only entropy flux $J_{(s)}$, energy flux $J_{(u)}$ and mass flux $J_{(\rho)}$. Then the temporal components of these four-vectors are related to each through Euler equation, i.e.
\be
J_{(u)}^0 = T J_{(s)}^0 + m J_{(\rho)}^0
\ee
or
\be
J_{(s)}^0 = \frac{1}{T} J_{(u)}^0 - \frac{m}{T} J_{(\rho)}^0 \label{eq:first}
\ee
and the spatial components through entropy current
\be
J_{(s)}^i = \frac{1}{T} J_{(u)}^i - \frac{m}{T} J_{(\rho)}^i \label{eq:current}
\ee
where $T$ is temperature and $m$ is chemical potential.  Equations \eqref{eq:first} and  \eqref{eq:current} be combined into a single equation for four-vectors
\be
J_{(s)}^\mu = \frac{1}{T} J_{(u)}^\mu - \frac{m}{T} J_{(\rho)}^\mu \label{eq:fout_current}
\ee
which together with conservation equations
\bea
\partial_\mu J_{(u)}^\mu &=& 0\\
\partial_\mu J_{(\rho)}^\mu &=& 0\\
\partial_\mu J_{(s)}^\mu &\equiv& R
\eea
implies an equation for entropy production
\be
R   = J_{(u)}^\mu \partial_\mu \left ( \frac{1}{T} \right ) +  J_{(\rho)}^\mu\partial_\mu \left (- \frac{m}{T} \right ).
\ee
The key (phenomenological) step is to assume that the fluxes can be expanded as 
\bea
J_{(u)}^\mu =  L_{11} \partial^\mu \left ( \frac{1}{T} \right ) +  L_{12}\partial^\mu \left (- \frac{m}{T} \right )\\
J_{(\rho)}^\mu =  L_{21} \partial^\mu \left ( \frac{1}{T} \right ) +  L_{22}\partial^\mu \left (- \frac{m}{T} \right )
\eea
where $L_{ij}$ is a two-by-two tensor which was shown by Onsager to be symmetric \cite{Onsager}. 

By following the same logic we expand the entropy production around equilibrium, i.e.
\be
\frac{1}{2\kappa}{\mR} = \mg_{\alpha\beta, \mu} \mJ^{\mu\alpha\beta} 
\ee
where the generalized forces are taken to be
\be
\mg_{\alpha\beta, \mu} \equiv \frac{\partial \mg_{\alpha\beta} }{\partial x^\nu} 
\ee
and fluxes are denoted by $\mJ^{\mu\alpha\beta}$. Since $ \mg^{\alpha\beta}$ is conjugate to information it is closely related to temperature. In the simplest case considered above the generalized forces are given by gradients of inverse temperature (or gradients of chemical potential times inverse temperature) and so it makes sense to treat partial derivatives of $\mg_{\alpha\beta}$ (or linear combinations of them) as generalized forces. Then we can expand fluxes around local equilibrium (i.e. when they vanish) to the linear order in generalized forces
\be
\mJ^{\mu\alpha\beta} =  \mL^{\mu\nu\;\alpha\beta\;\gamma\delta} \mg_{\gamma\delta, \nu }.
\ee
and thus
\be
\frac{1}{2\kappa}{\mR} = \mL^{\mu\nu\;\alpha\beta\;\gamma\delta} \mg_{\alpha\beta, \mu}  \mg_{\gamma\delta, \nu } \label{eq:Onsager2}.
\ee
In fact one can think of \eqref{eq:Onsager2} as a defining equation for the Onsager tensor, but then we are forced to only consider Onsager tensors $ \mL^{\mu\nu\;\alpha\beta\;\gamma\delta}$ that are symmetric under exchange  $(\mu, \alpha,\beta) \leftrightarrow (\nu, \gamma, \delta)$, i.e.
 \be
 \mL^{\mu\nu\;\alpha\beta\;\gamma\delta} = \mL^{\nu\mu\;\beta\alpha\;\delta\gamma}.  \label{eq:onsager_relation}
 \ee
These are the Onsager reciprocity relations \cite{Onsager} derived for our computational system. 

For example, consider a tensor
\be
\mL^{\mu\nu\;\alpha\beta\;\gamma\delta} = \frac{1}{2\kappa} \left (  \mg^{\alpha\nu} \mg^{\beta\delta} \mg^{\mu\gamma} +\mg^{\alpha\gamma} \mg^{\beta\nu} \mg^{\mu\delta} -\mg^{\alpha\gamma} \mg^{\beta\delta} \mg^{\mu\nu}  \right ) \label{eq:simple_choice}
\ee
for which the flux can be rewritten as 
\be
\mJ^{\mu\alpha\beta} = \frac{1}{\kappa}\mg^{\alpha\gamma} \mg^{\beta\delta} \Gamma^{\mu}_{\phantom{\mu}\gamma\delta}
\ee
 where
 \be
 \Gamma^{\mu}_{\phantom{\mu}\gamma\delta} \equiv \frac{1}{2} \mg^{\mu\nu} \left ( \mg_{\nu\gamma,\delta} +  \mg_{\nu\delta,\gamma} -  \mg_{\gamma\delta, \nu} \right )
 \ee
 are the Christoffel symbols and $\kappa$ is some phenomenological constant which measures the strength of the emergent gravitational effects. If we inset it back into the entropy functional we get
 \bea
 \int d^{D+1} x \sqrt{|\mg|} \frac{1}{2\kappa}{\mR} &=&\frac{1}{\kappa}\int d^{D+1} x \sqrt{|\mg|} \mg_{\alpha\beta, \mu} \mg^{\alpha\gamma} \mg^{\beta\delta} \Gamma^{\mu}_{\phantom{\mu}\gamma\delta} \notag  \\
 &=&- \frac{1}{\kappa} \int d^{D+1} x \sqrt{|\mg|} \mg^{\gamma\delta}_{\phantom{\alpha\gamma},\mu}  \Gamma^{\mu}_{\phantom{\mu}\gamma\delta} \notag\\
  &=& \frac{1}{\kappa}\int d^{D+1} x \sqrt{|\mg|} \left ( \mg^{\alpha\gamma}_{\phantom{\alpha\gamma}}  \frac{\partial}{\partial x^\mu} \Gamma^{\mu}_{\phantom{\mu}\gamma\alpha} +   \mg^{\alpha\gamma} \Gamma^{\mu}_{\phantom{\mu}\gamma\alpha} \frac{1}{\sqrt{|\mg|}}  \frac{\partial}{\partial x^\mu} \left ( \sqrt{|\mg|} \right )\right ) \notag\\
  &=&\frac{1}{\kappa}\int d^{D+1} x \sqrt{|\mg|} \left ( \mg^{\alpha\gamma}_{\phantom{\alpha\gamma}}  \frac{\partial}{\partial x^\mu} \Gamma^{\mu}_{\phantom{\mu}\gamma\alpha} +   \mg^{\alpha\gamma} \Gamma^{\mu}_{\phantom{\mu}\gamma\alpha} \Gamma^\nu_{\phantom{\nu}\mu\nu}\right ) \notag\\
  &=&\frac{1}{\kappa} \int d^{D+1} x \sqrt{|\mg|} \mg^{\mu\nu} \left (\Gamma^{\alpha}_{\phantom{\alpha}\mu\nu, \alpha} +  \Gamma^{\beta}_{\phantom{\beta}\mu\nu} \Gamma^\alpha_{\phantom{\alpha}\alpha\beta}\right )
 \eea
 where
 \be
 \Gamma^\alpha_{\phantom{\alpha}\mu\nu, \beta} \equiv \frac{\partial}{\partial x^\beta}  \Gamma^\alpha_{\phantom{\alpha}\mu\nu}.
 \ee
 However, this action is not covariant unless we anti-symmetrize it with respect to lower indices $\alpha$ and $\mu$ which would produce the Einstein-Hilbert action, i.e.
\be
\int d^{D+1} x \sqrt{|\mg|} \frac{1}{2\kappa}{\mR}  =\frac{1}{\kappa}  \int d^{D+1} x \sqrt{|\mg|} \mg^{\mu\nu}  \left (\Gamma^{\alpha}_{\phantom{\alpha}\nu[\mu, \alpha]} +  \Gamma^{\beta}_{\phantom{\beta}\nu[\mu} \Gamma^\alpha_{\phantom{\alpha}\alpha]\beta}\right )  
  \label{eq:einstein_hilbert}
\ee
where the anti-symmetrization (and also symmetrization) was defined in Eq. \eqref{eq:symmetrization}.

From Onsager's relations \eqref{eq:onsager_relation} we know that $ \mL^{\mu\nu\;\alpha\beta\;\gamma\delta} $ should be symmetric under exchange of indices  $(\mu, \alpha,\beta) \leftrightarrow (\nu, \gamma, \delta)$, but there are also other (trivial) symmetries that one should impose $(\alpha) \leftrightarrow (\beta)$, $(\gamma) \leftrightarrow (\delta)$, due to symmetries of the metric, i.e.
\be
 \mL^{\mu\nu\;\alpha\beta\;\gamma\delta}  =  \mL^{\mu\nu\;(\alpha\beta)\;(\gamma\delta)} \label{eq:correct_choice} 
\ee
The overall space of such tensors is still pretty large and we would not be able to proceed further without making additional assumptions, but it turns out that a very simple choice leads to general relativity, i.e.
\be
\mL^{\mu\nu\;\alpha\beta\;\gamma\delta} =  \frac{1}{8\kappa} \left (  \mg^{\alpha\nu} \mg^{\beta\delta} \mg^{\mu\gamma} +\mg^{\alpha\gamma} \mg^{\beta\nu} \mg^{\mu\delta} -\mg^{\alpha\gamma} \mg^{\beta\delta} \mg^{\mu\nu} -\mg^{\alpha\beta} \mg^{\gamma\delta} \mg^{\mu\nu} \right ) .\label{eq:Onsager}
\ee
After integrating by parts, neglecting the boundary terms and collecting all other terms we get
\bea
\int d^{D+1} x &\sqrt{|\mg|} &\frac{1}{2\kappa}{\mR}  =
  \int d^{D+1} x \sqrt{|\mg|} \mg^{\mu\nu} \;\; \frac{1}{\kappa} \left (\Gamma^{\alpha}_{\phantom{\alpha}\nu[\mu, \alpha]} +  \Gamma^{\beta}_{\phantom{\beta}\nu[\mu} \Gamma^\alpha_{\phantom{\alpha}\alpha]\beta}\right )=\\\notag  
  =&\int& d^{D+1} x \sqrt{|\mg|} \;\; \frac{1}{8\kappa} \left (  \mg^{\alpha\nu} \mg^{\beta\delta} \mg^{\mu\gamma} +\mg^{\alpha\gamma} \mg^{\beta\nu} \mg^{\mu\delta} -\mg^{\alpha\gamma} \mg^{\beta\delta} \mg^{\mu\nu} -\mg^{\alpha\beta} \mg^{\gamma\delta} \mg^{\mu\nu} \right )  \mg_{\alpha\beta,\mu} \mg_{\gamma\delta,\nu}\;. 
\eea
Upon varying the full action \eqref{eq:full_action} with respect to the metric (which corresponds to extremization of entropy production) we arrive at the Einstein equations
\be
\mR_{\mu\nu} - \frac{1}{2} \mR \mg_{\mu\nu} + \Lambda \mg_{\mu\nu} = \kappa \left \langle T_{\mu\nu} \right \rangle
\ee
where
\be
\mR_{\mu\nu} \equiv 2 \left (\Gamma^{\alpha}_{\phantom{\alpha}\nu[\mu, \alpha]} +  \Gamma^{\beta}_{\phantom{\beta}\nu[\mu} \Gamma^\alpha_{\phantom{\alpha}\alpha]\beta}\right )
\ee
is the Ricci tensor.  Of course, the expectations are that this result would only hold near equilibrium, and there should be deviations from general relativity when some of the symmetries in the Onsager tensor are broken. 

What is less clear is why (or when) this particular choice of the Onsager tensor \eqref{eq:Onsager} is preferred when it comes to a non-equilibrium dynamics of the metric. As of now, we only know that \eqref{eq:correct_choice} must be satisfied, but \eqref{eq:Onsager} has a lot more symmetries and as a result of these symmetries we are led to a covariant theory of general relativity described by the entropy production which has the form of the Einstein-Hilbert action. To this end, it would be interesting to study if one is forced to impose additional restrictions on the computational dynamics of wave-functions or if these symmetries are generically expected closer to equilibrium. We leave all these questions for future research.

\section{Conclusion}\label{sec:discussion}

In this paper we achieved two main results: constructed a fully covariant information (and also computation) theory (Secs. \ref{sec:ensembles}, \ref{sec:computation})  and then developed (equilibrium and non-equilibrium) thermodynamic description of information (Secs. \ref{sec:thermodynamics}, \ref{sec:gravity}) for either statistical or quantum systems. Although the first result motivated the development of the second, it is important to note that these two results can be used independently. In fact, the situation is not very different from the usual (effective) field theories that are useful at small temperatures and thermodynamic models that are useful at high temperatures. 

In our case the role of temperature is played by the (inverse) metric tensor and so if the quantities, such as curvature, remain small we expect that the covariant information theory would provide a good description of the dynamics of information. However, when the curvature becomes large, its own dynamics cannot be ignored and the covariant information theory description becomes less useful. In this regime one can get a lot more insight by considering a thermodynamic description that we started to develop in this paper. This does not mean that one should give up on the notion of effective fields, but these fields (such as metric) are no longer fundamental. In particular, we described how general relativity (which is a field theory) provides a good (but only approximate) description of non-equilibrium (thermo) dynamics of the metric for a particularly simple and highly symmetric form of the Onsager tensor. But further away from equilibrium it is expected that some of the symmetries of the Onsager tensor would be broken which should give rise to deviations from general relativity. 

Let us also remark that, although the emergent dynamics of the metric tensor that we derived was obtained as a thermodynamic limit of some abstract computational system it might very well be the way the physical gravity that we observe around us really works. Of course, we are not even close to showing that, since the only matter field that we had in our dual field theory was the scalar infoton field, but one can also add an auxiliary Abelian fields as in Ref. \cite{Vanchurin}. However, what one really wants is to understand which computational system gives rise not just to general relativity, but to all other fields of the standard model and also to dark matter, dark energy and perhaps cosmic inflation. From this perspective the proposed theory should be considered as a toy-model in which nevertheless the gravity did emerged from a purely computational system. And this allows us to start asking other questions related to the physical world in context of the non-equilibrium thermodynamics of information.  For example, it would be interesting to see if one can attribute the breaking of the Onsager symmetries to, for example, dark matter. Moreover, we recall that in equation \eqref{eq:full_action} in addition to the Einstein-Hilbert term we also get a cosmological constant term which is nothing but integration constant of the entropy functional that first appeared in \eqref{eq:action2}. It would be interesting to see if this integration constant (which has thermodynamic origin) could be responsible for the observed accelerated expansion of the Universe or dark energy. Moreover, if general relativity is indeed a non-equilibrium process then it would be interesting to analyze other outstanding cosmological problems (e.g. entropy problem, measure problem, etc.) perhaps along the lines of ideas described in Ref. \cite{dynamical_systems} where it was also argued that the (higher order than Onsager) symmetries described by the fluctuation-dissipation theorem might be observable in the cosmic microwave background radiation.

{\it Acknowledgments.} The author is grateful to Alan Guth, Mudit Jain, Thomas Jordan, Mahdiyar Noorbala, Daniel Schubring and Evan Severson for very useful discussions. The work was supported in part by  Foundational Questions Institute (FQXi).

\end{document}